\documentclass{PoS}

\usepackage{graphicx,epsfig,subfigure,rotating}
\usepackage{fontenc, times, mathptmx}

\usepackage{amssymb,amsmath}

\title{Fundamental limits of radio interferometers: calibration and source parameter estimation}

\ShortTitle{Fundamental limits of radio interferometers: calibration and source parameter estimation}

\author{\speaker{Cathryn Trott}%
         \thanks{ARC Centre of Excellence for All-Sky Astrophysics (CAASTRO)}\\
        ICRAR/Curtin University/CAASTRO\\
        E-mail: \email{cathryn.trott@curtin.edu.au}}

\author{Randall Wayth\\
        ICRAR/Curtin University\\
        E-mail: \email{r.wayth@curtin.edu.au}}

\author{Steven Tingay\\
        ICRAR/Curtin University\\
        E-mail: \email{s.tingay@curtin.edu.au}}

\abstract{We use information theory to derive fundamental limits on the capacity to calibrate next-generation radio interferometers, and measure parameters of point sources for instrument calibration, point source subtraction, and data deconvolution. We demonstrate the implications of these fundamental limits, with particular reference to estimation of the 21cm Epoch of Reionization power spectrum with next-generation low-frequency instruments (e.g., the Murchison Widefield Array -- MWA, Precision Array for Probing the Epoch of Reionization -- PAPER), where short time scale instrumental calibration is required due to the impact of the ionosphere on the signal wavefront. Finally, we explore the optimal point source precision available by using a combination of current and prior information. Estimation schemes that incorporate prior information may be advantageous when the measurement precision is comparable to the characteristic refraction scale of the ionosphere.}

\FullConference{Resolving The Sky - Radio Interferometry: Past, Present and Future \\
                 April 17-20,2012\\
                 Manchester, UK}

\begin{document}

\section{Introduction}
The suite of radio interferometers currently under construction and in planning phases, has the potential to deliver answers to exciting scientific questions, such as describing the properties and evolution of early universe structures, and as a probe of variable radio sources. With this new era of radio interferometry comes the development of new instruments and technologies, as well as the challenges inherent in advancing a field and gaining new knowledge. For most programs, the sensitivity and resolution limits of our instruments will be pushed, and there will be an increasing reliance on temporal observing modes and the low-frequency domain. We will see a shift from narrow-field, long integration, few receiving-element science, to wide-field, snapshot observing with large element arrays, culminating in the ultimate multi-science instrument: the Square Kilometre Array (SKA). In addition to the engineering challenges inherent in the design and execution of the SKA, there are accompanying questions to be answered regarding the data limitations for such an instrument, its calibratability and realisable dynamic range. These questions are being increasingly addressed in the literature, and suggest a complicated path forward for the optimal design of the instrument and our data analysis methodology \cite{wijnholds08}.

Key science goals of the SKA, and pathfinder instruments, demand precision \textit{quantitative} radio astronomy, where an understanding of the underlying properties of the data is crucial for robust and unbiased science results. Such goals include the detection of new classes of fast transient sources, and detection and estimation of the neutral hydrogen signal from the Epoch of Reionization (EoR). These goals are shared by many instruments currently in the development, building and commissioning phases, such as the Murchison Widefield Array (MWA), Precision Array for Probing the Epoch of Reionization (PAPER), Low Frequency Array (LOFAR), and the Long Wavelength Array (LWA) \cite{lonsdale09,tingay12,parsons10,stappers11,ellingson09}. These instruments and projects operate at low frequencies (100-200 MHz), where the ionopshere plays an important role in the shape of the signal wavefront.

The impact of the ionosphere on the measured correlations (visibilities) ranges from zero (constant phase shift between antennas) to de-focussing, depending on the extent of the array, the frequency, and the characteristic size scale of ionospheric disturbances.
Cohen \& R{\"o}ttgering \cite{cohen09} studied the differential refraction (position change relative to other sources in the field) for Very Large Array (VLA) fields at 74~MHz, yielding a source shift of $\sim$100 arcseconds at a separation of 25 degrees under normal ionospheric conditions.

Wide-field low frequency instruments, such as the MWA, PAPER and LWA, aim to use known bright point sources as calibrators to model the instantaneous effect of the ionosphere on the wavefront. In this work we consider an observational scheme where the calibration is performed in real-time, necessitating a measurement of the instrument and sky response on eight to ten second timescales. These solutions are applied in real-time to the measured data (visibilities).

The precision with which the sky position of calibration sources may be estimated with limited data impacts the degree to which an instrument is calibratable, and consequently on the quality of scientifically-relevant metrics. We begin by deriving the theoretical precision with which point source parameters can be estimated from a given visibility dataset, and propagate these errors to additional noise terms in the visibilities. As an example scientific application of this residual signal, we then propagate the errors to a metric of interest for the statistical estimation of the 21~cm EoR signal: the angular power spectrum. We then discuss estimation whereby prior information is balanced with information from the current dataset to estimate parameters, and demonstrate the point source position precision that may be achieved by an optimal estimator.

\section{Point source estimation limits}
\subsection{The Cramer-Rao bound (CRB)}
We use the Cramer-Rao lower bound (CRB) on the precision of parameter estimates. The CRB calculates the precision with which a minimum-variance unbiased estimator could estimate a parameter value, \textit{using the information content of the dataset}. It is computed as the square-root of the corresponding diagonal element of the inverse of the Fisher information matrix (FIM). The ($ij$)th entry of the FIM for a vector ${\boldsymbol{\theta}}$ of unknown parameters is given by \cite{kay93}:
\begin{equation}
[\boldsymbol{I(\theta)}]_{ij} = -E\left[\frac{\partial^2{\ln{L({\bf{x}};{\boldsymbol{\theta}})}}}{\partial{\theta_i}\partial{\theta_j}} \right],
\end{equation}
where $L$ denotes the likelihood function describing the likelihood of measuring a dataset, ${\bf x}$ for a given parameter set, $\boldsymbol{\theta}$. The CRB places a fundamental lower limit on the measurement precision of any parameter. In this work it will be used to gain an understanding of the fundamental limits of point source subtraction, and how these impact EoR estimation.

\subsubsection{Incorporating prior information -- the hybrid CRB}\label{hcrlb}
If one uses some prior information about the value of a parameter, in addition to the information contained within the current dataset, one can use a hybrid CRB (HCRLB) to obtain a measure of the maximum estimation precision. In this case, the maximal precision corresponds to the minimum mean squared error, and may be biased by the use of inaccurate prior information. We extend the formalism to incorporate the parameter of interest in the likelihood function, treating the parameter as random rather than deterministic. Using Bayes' rule, and considering the joint likelihood for the data and the parameter with prior information, $\theta$:
\begin{equation}
L({\bf x},\theta) = L({\bf x}|\theta)L(\theta),
\end{equation}
where $L({\bf x}|\theta)$ is the likelihood of the data conditioned on the parameter value. The FIM is;
\begin{equation}
[\boldsymbol{I(\theta)}]_{ij} = -E\left[\frac{\partial^2{\ln{L({\bf{x}}|{\theta})}}}{\partial{\theta_i}\partial{\theta_j}} \right] -E\left[\frac{\partial^2{\ln{L(\theta)}}}{\partial^2{\theta}} \right]. \label{prior_eqn}
\end{equation}
The final term in equation \ref{prior_eqn} contains the prior information available about the value of parameter $\theta$. For a Gaussian-distributed likelihood function, the Fisher information contains an additional term corresponding to the prior information, given by,
$I_{\theta\theta} = 1/\sigma^2$,
where $\sigma^2$ is the variance of the Gaussian distribution. A smaller variance corresponds to precise prior information, and contributes more information to the Fisher matrix than a broad distribution. In reference to this work, we are considering the effect of the ionosphere to produce a Gaussian-distributed shift in the position of a source, with a characteristic refraction scale dependent on the level of ionospheric activity on the timescale of interest ($\sigma$).

\subsection{Residual signal in visibilities}
\begin{figure}
\begin{center}
\subfigure[MWA -- hypothetical uniform $uv$ coverage.]{\includegraphics[scale=0.3]{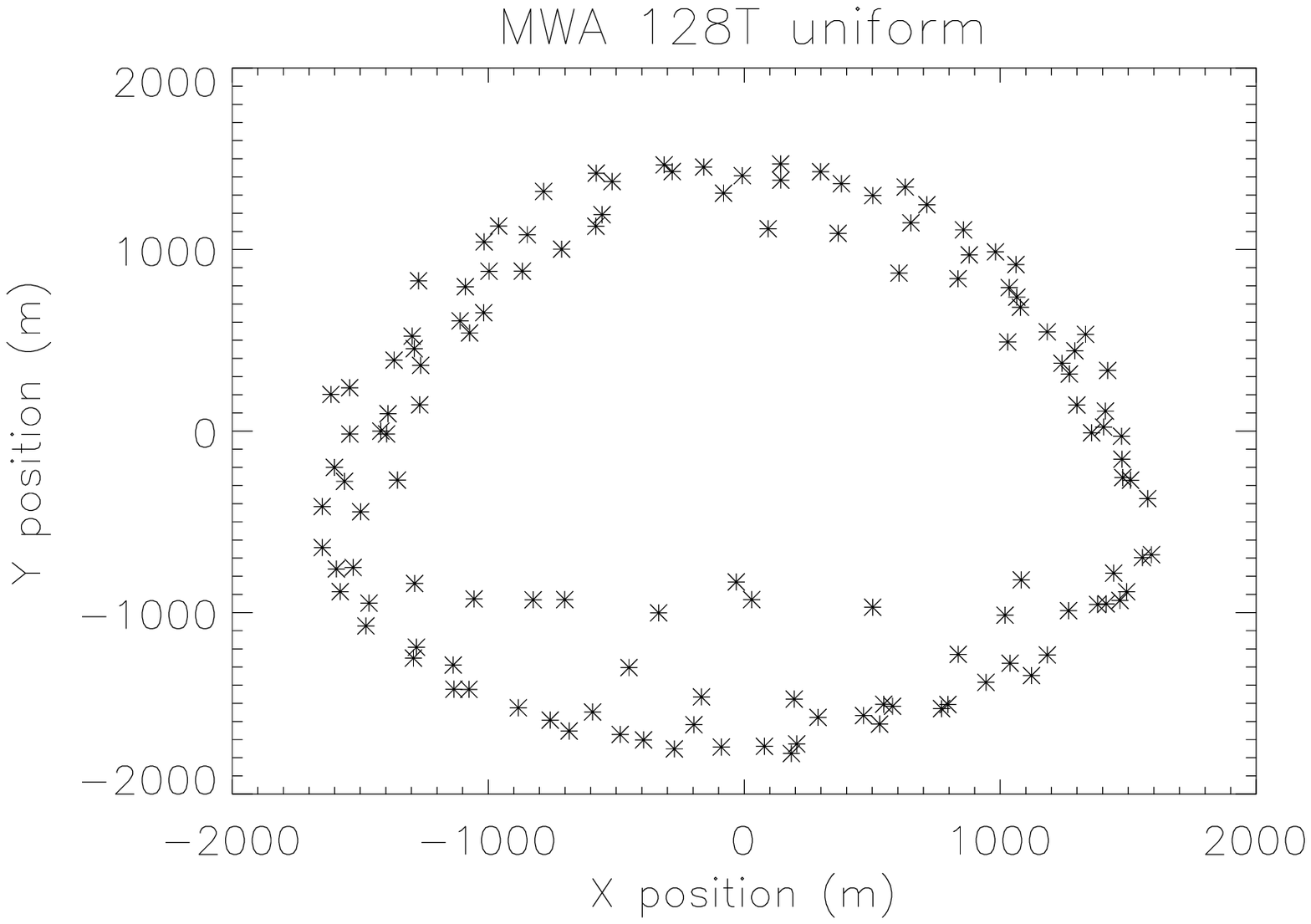}}
\subfigure[MWA.]{\includegraphics[scale=0.3]{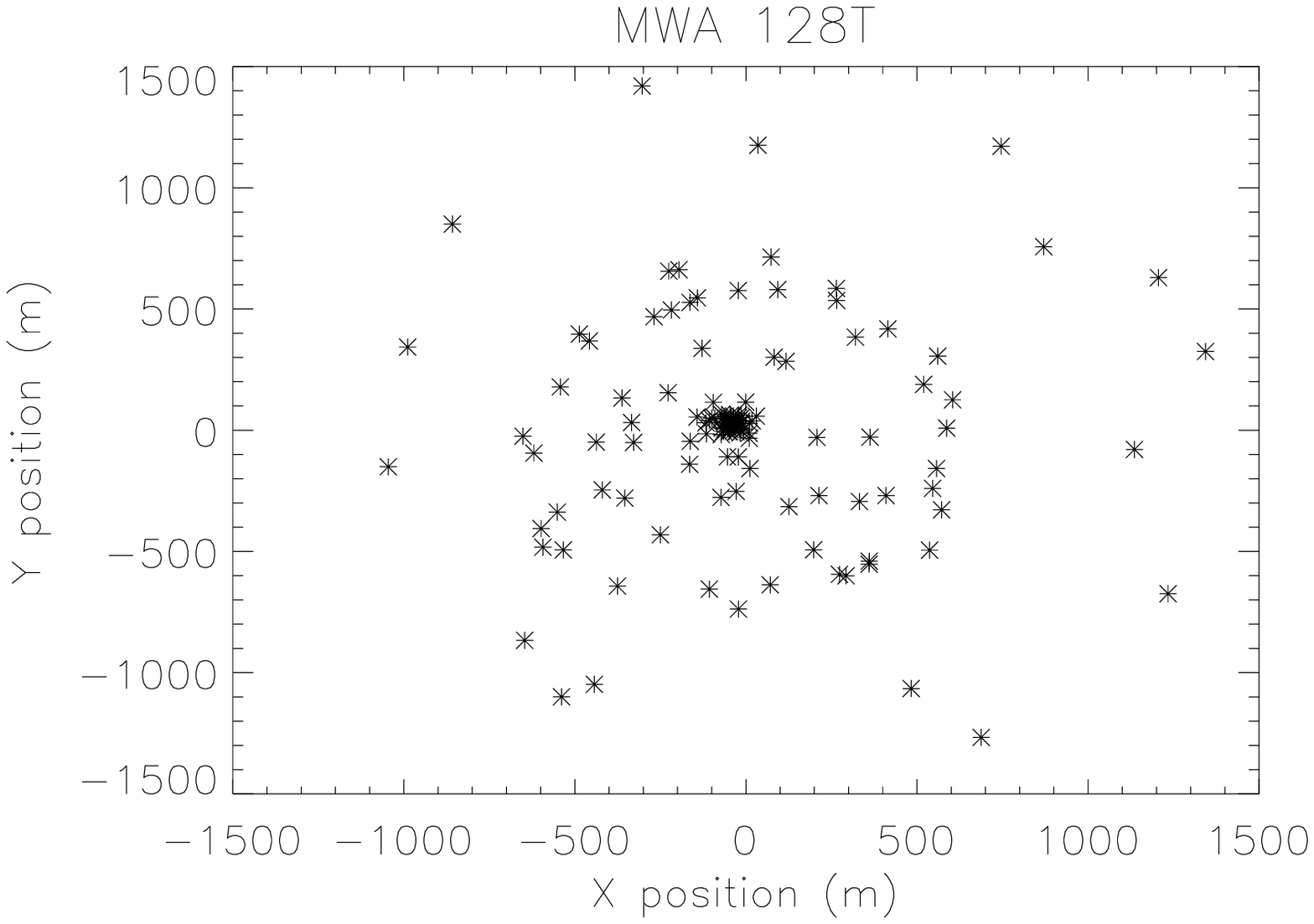}}\\
\subfigure[Actual 64-dipole minimum-redundancy (PAPER).]{\includegraphics[scale=0.3]{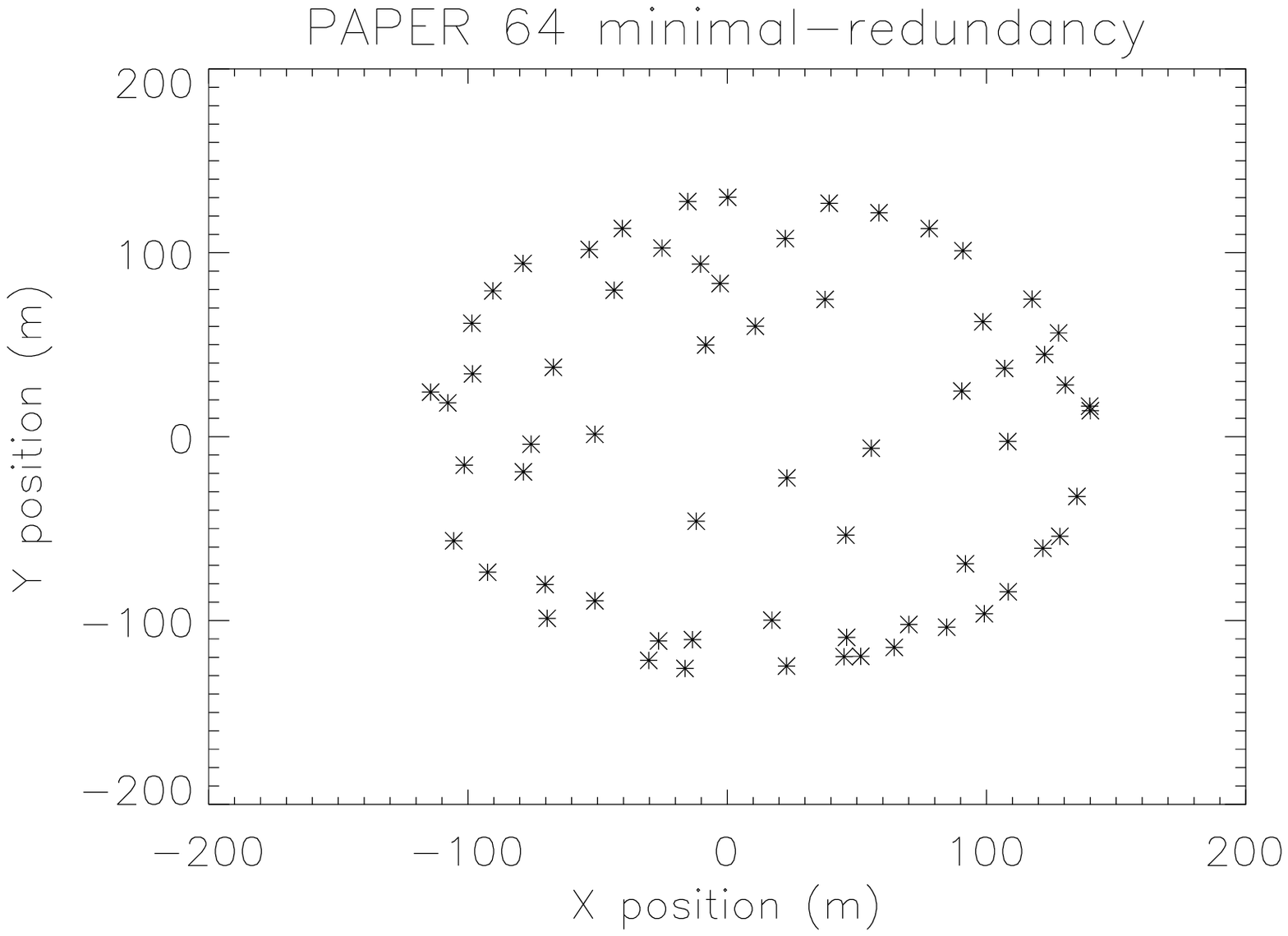}}
\subfigure[Potential 64-dipole maximum-redundancy (PAPER).]{\includegraphics[scale=0.3]{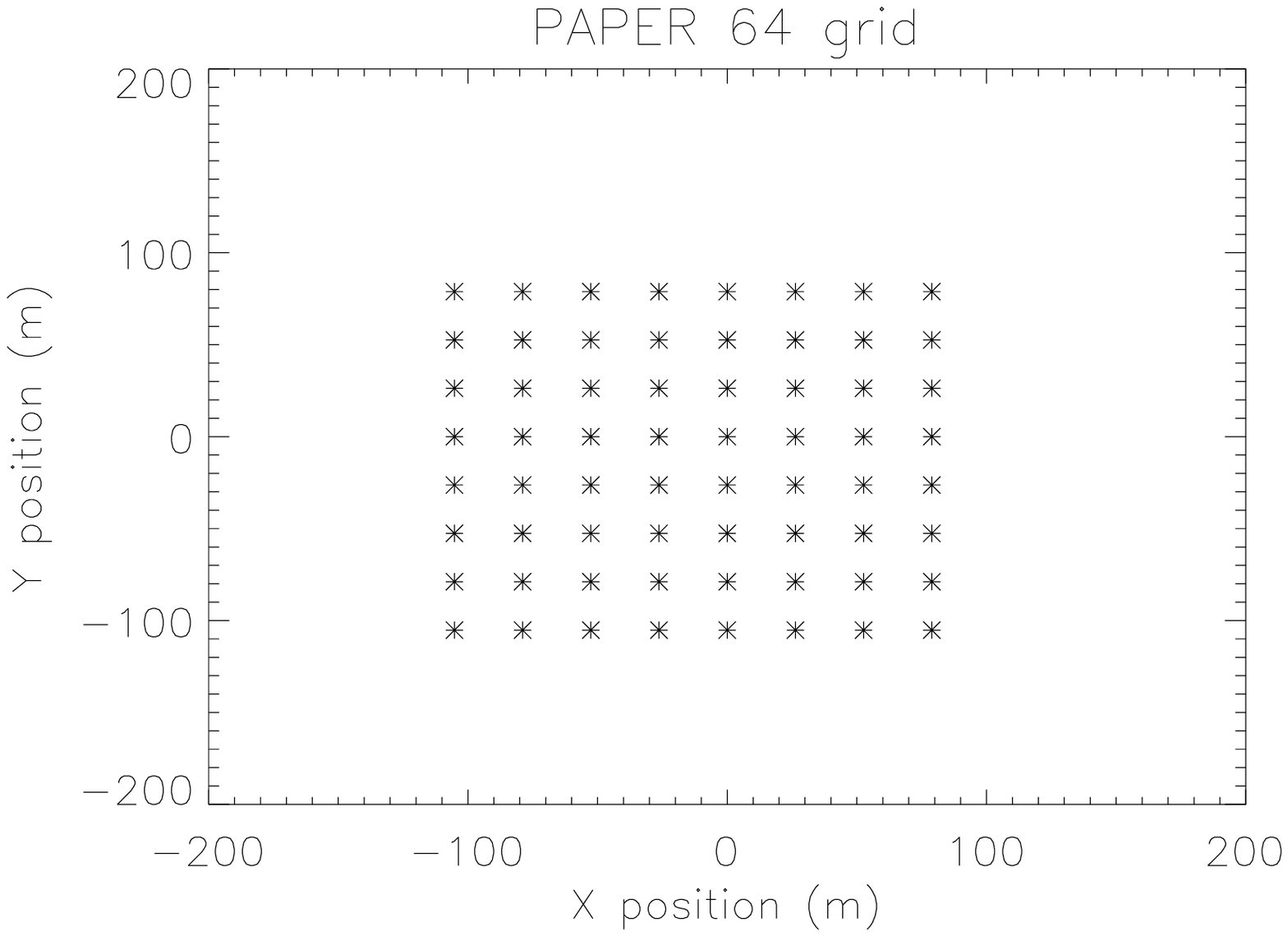}}
\caption{Antenna configurations for the MWA and PAPER considered in this work.}
\label{antenna_config}
\end{center}
\end{figure}

\begin{figure}
\subfigure[Dataset alone.]{\includegraphics[scale=0.4]{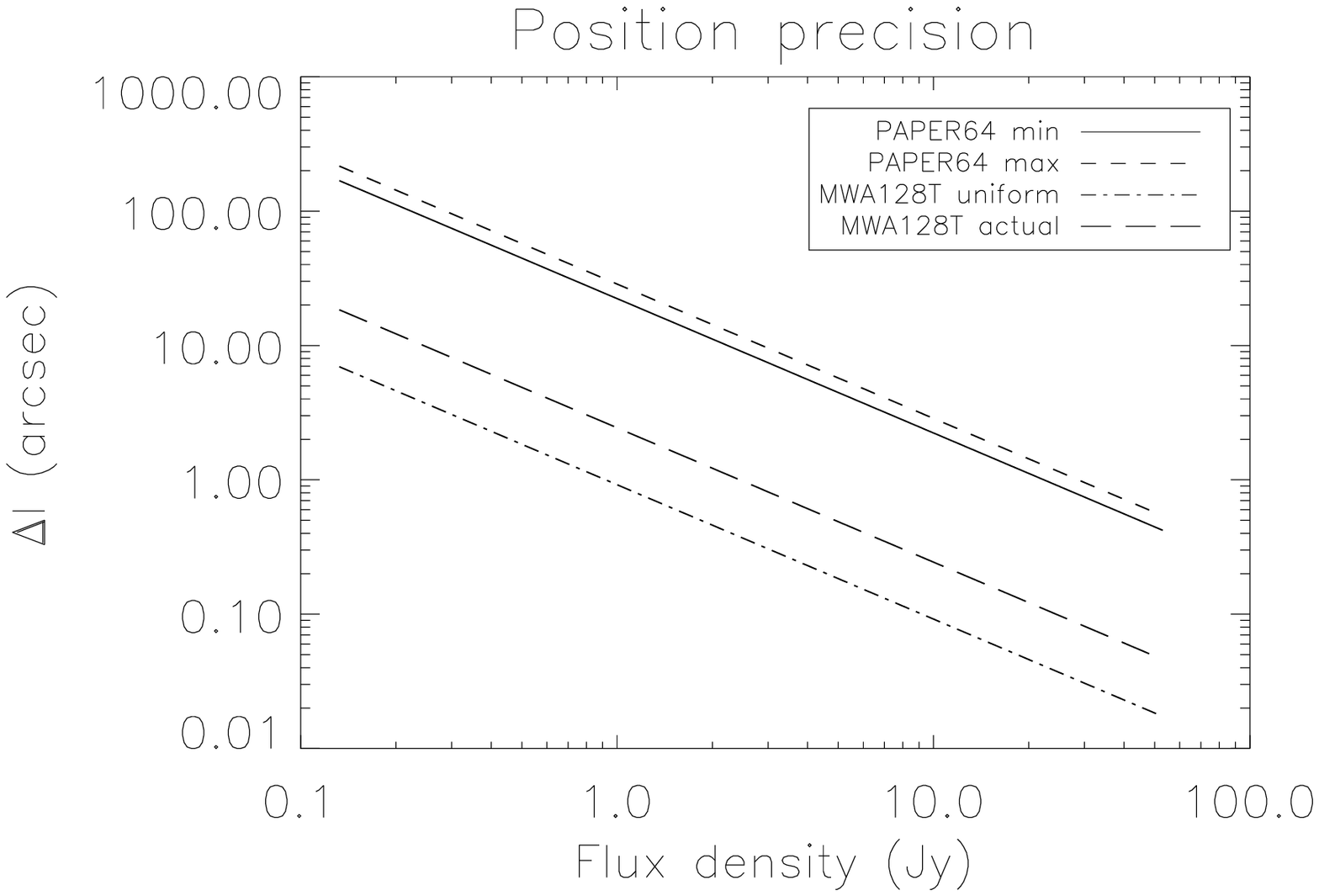}}
\subfigure[Characteristic ionospheric refraction scale: 60".]{\includegraphics[scale=0.4]{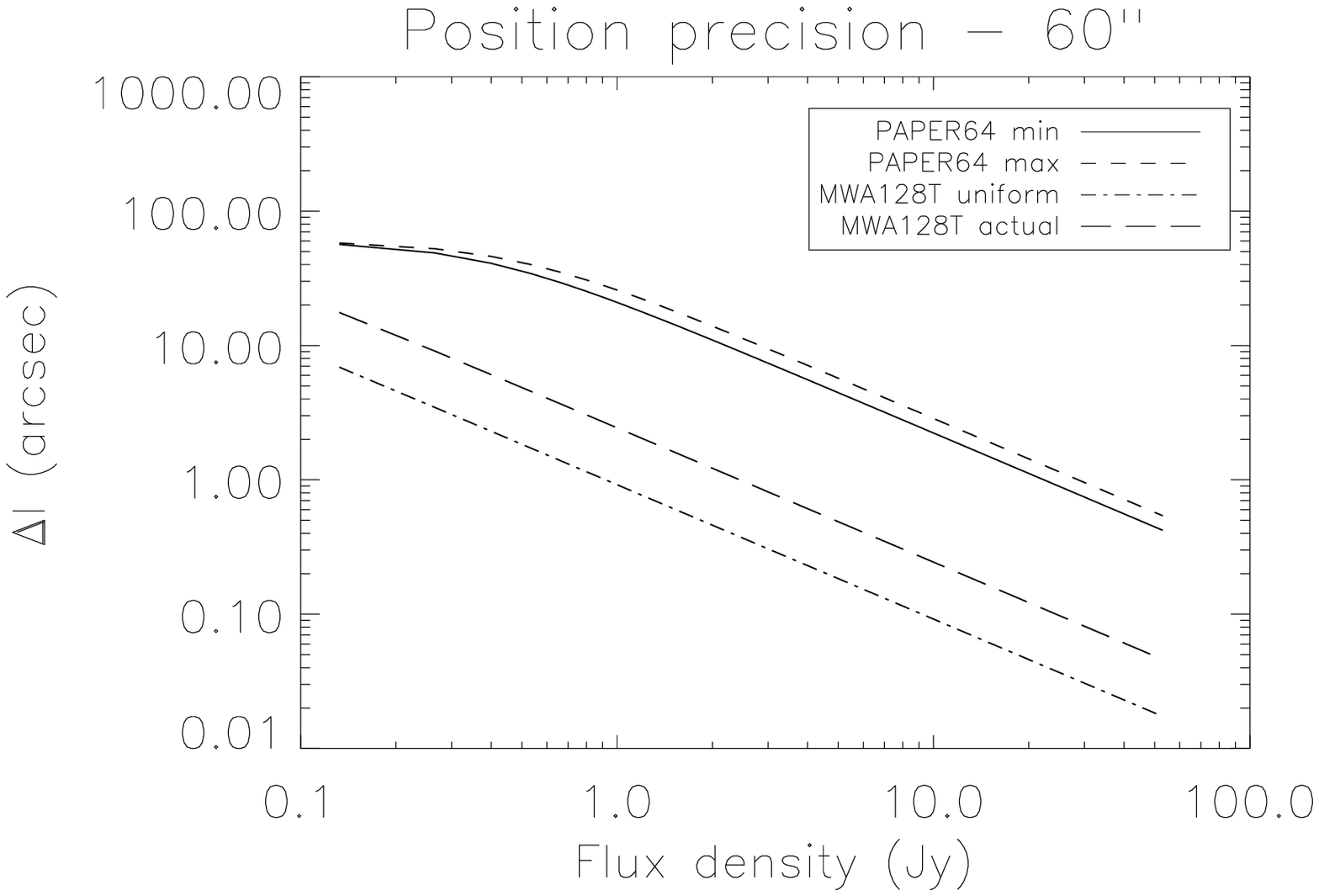}}
\subfigure[Characteristic ionospheric refraction scale: 10".]{\includegraphics[scale=0.4]{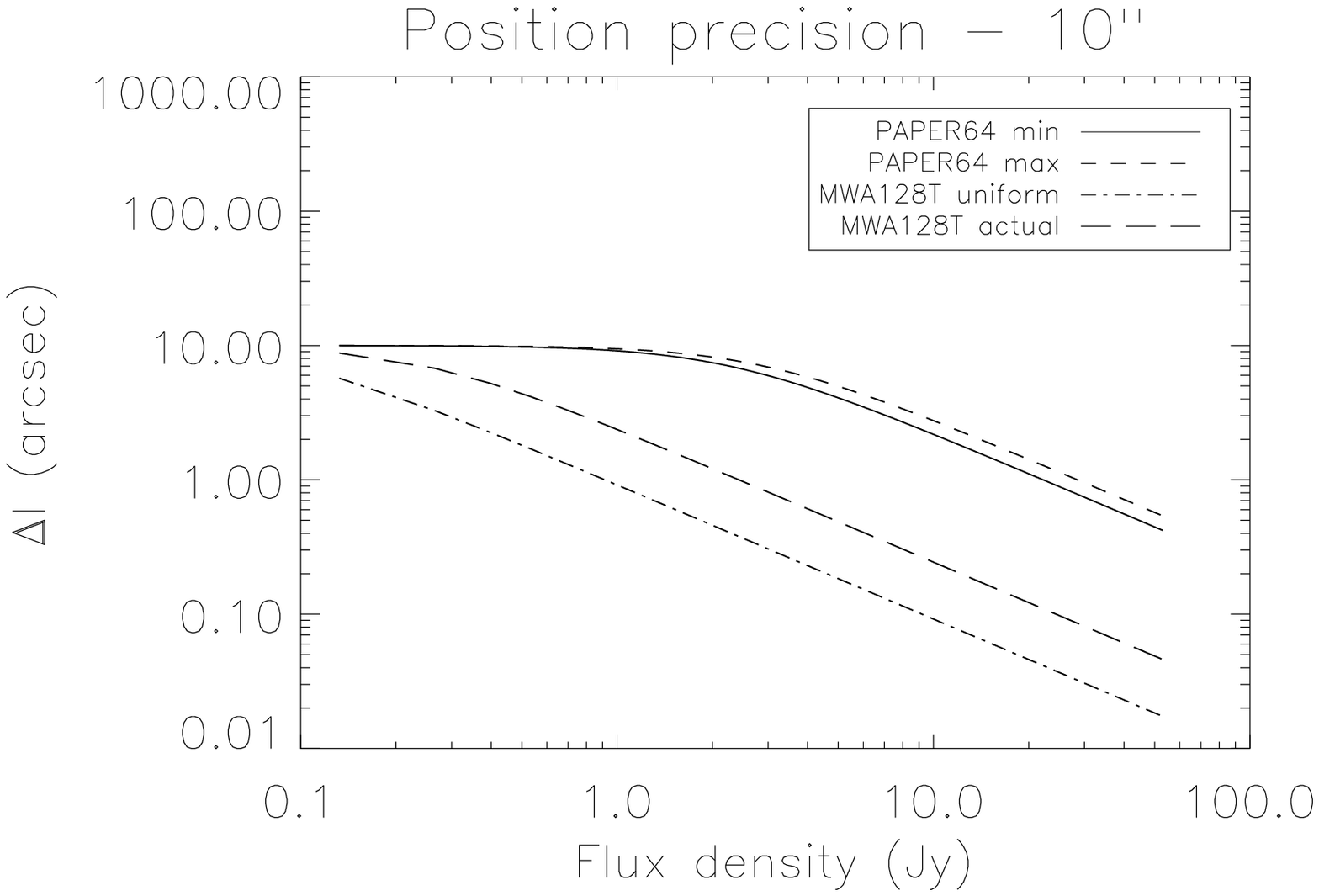}}
\caption{Optimal source sky position precision (minimum mean squared error) for the dataset alone (a), and different characteristic ionospheric refraction scales (b, c).}
\label{hcrlb_precision}
\end{figure}

The signal in each visibility is the linear combination of signals from each calibrator, and is given by:
\begin{equation}
\tilde{s}[f,n] = \displaystyle\sum_{i=1}^{N_c} V_i(u_{fn},v_{fn}) = \displaystyle\sum_{i=1}^{N_c} B_i(l_i,m_i)\exp{\left[-2{\pi}i(u_{fn}l_i+v_{fn}m_i)\right]},
\label{visibility_signal}
\end{equation}
where $N_c$ is the total number of calibrators, described by source strength, $B_i$, located at sky position, ($l_i,m_i$), for a baseline, $n$, and frequency channel, $f$. The signal is embedded within white Gaussian thermal noise, with diagonal covariance matrix, $\boldsymbol{C}=\sigma^2\boldsymbol{I}$, where $\sigma$ is set by the radiometer noise.
Using the full dataset, and \textit{no prior information}, the minimum uncertainty in the parameter estimates for calibrator, $i$, and their non-zero covariances, are given by \cite{trott11}:
\begin{eqnarray}
\Delta{l_i} &\geq& \frac{\sigma I_{v^2}^{1/2}}{2\sqrt{2}\pi{B_i}} 
\left[ I_{v^2} I_{u^2} - \left(I_{uv} \right)^2 
\right]^{-1/2} \label{l_precision}\\
\Delta{m_i} &\geq& \frac{\sigma I_{u^2}^{1/2}}{2\sqrt{2}\pi{B_i}} 
\left[ I_{v^2} I_{u^2} - \left(I_{uv}\right)^2 \right]^{-1/2} \label{m_precision}\\
\Delta{B_i} &\geq& \frac{\sigma}{\sqrt{2NF}} \\ 
{\rm cov}(l_i,m_i) &=& -\frac{\sigma^2 I_{uv}}{8\pi^2{B_i}^2} 
\left[ I_{v^2} I_{u^2} - \left(I_{uv}\right)^2 \right]^{-1}
\end{eqnarray}
 where
\begin{equation}
I_{u^2} = \displaystyle\sum_{n=1}^N \displaystyle\sum_{f=1}^F  u_{fn}^2,  \hspace{4mm}
I_{uv} = \displaystyle\sum_{n=1}^N \displaystyle\sum_{f=1}^F  u_{fn}v_{fn},  \hspace{4mm}
I_{v^2} = \displaystyle\sum_{n=1}^N \displaystyle\sum_{f=1}^F v_{fn}^2. \label{Iv2}
\end{equation}
We perform this analysis for two upcoming statistical EoR experiments: the MWA and PAPER. Each array is modelled with two antenna configurations, corresponding to minimally-redundant and maximally-redundant $uv$ sampling. Figure \ref{antenna_config} displays the antenna configurations for the four arrays considered, and Table \ref{instrument_design} displays the experiment parameters.
\small
\begin{table}
\begin{center}
\begin{tabular}{|c|c|c|}
\hline Parameter & Value (MWA) & Value (PAPER) \\ 
\hline\hline N$_{\rm ant}$ & 128 & 64 \\ 
\hline $\nu_0$ & 150 MHz & 150 MHz \\
\hline Bandwidth & 30.72 MHz & 70.0 MHz \\
\hline $\Delta\nu$ & 125 kHz & 125 kHz \\
\hline Field-of-view & 30$^o$ & 60$^o$ \\ 
\hline Calibrators ($>$1 Jy) & 392 & 1579\\
\hline T$_{\rm sys}$ & 440K & 440K \\ 
\hline $\Delta{t}$ & 8s & 8s \\ 
\hline $t_{\rm tot}$ & 300h & 720h \\ 
\hline $d_{\rm max}$ & 3000m & 260m \\ 
\hline
\end{tabular}
\caption{Parameters used for the primary instrument design of the MWA and PAPER 64-dipole instruments.}\label{instrument_design}
\end{center}
\end{table}
\normalsize
Figure \ref{hcrlb_precision}(a) plots the optimal point source position precision ($\Delta{l}$) as a function of calibrator signal strength (Jy) for these four instruments.
The linear dependence of precision with signal strength is observed. The inner core $+$ outer ring structure of the MWA degrades its performance slightly compared with the uniform array. The short baselines and fewer antennas of PAPER degrade their ability to localize sources well compared with the MWA, although wider instantaneous bandwidth balances this effect somewhat.

\subsection{Propagation of errors into interferometric visibilities}\label{error_prop_section}
The uncertainties in signal parameters for each calibrator are propagated into uncertainty in the measured visibilities, in addition to the statistical thermal noise. The treatment provides a full covariant analysis, including any correlations between visibilities due to residual noise power. For a given visibility and a single calibrator, the variance is given by:
\small
\begin{eqnarray}
\sigma_V^2 &=& \left|\frac{\partial{V}}{\partial{l}} \right|^2 \sigma_l^2 + \left|\frac{\partial{V}}{\partial{m}} \right|^2 \sigma_m^2 + 2\left(\frac{\partial{V}^\ast}{\partial{l}}\frac{\partial{V}}{\partial{m}}+\frac{\partial{V}}{\partial{l}}\frac{\partial{V}^\ast}{\partial{m}} \right) {\rm cov}_{lm} + \left|\frac{\partial{V}}{\partial{B}} \right|^2 \sigma_B^2\label{error_vis1}\\
&=& 4\pi^2u^2B^2\frac{\sigma^2}{8\pi^2B^2}f_1(u,v) + 4\pi^2v^2B^2\frac{\sigma^2}{8\pi^2B^2}f_2(u,v) + 8\pi^2uvB^2\frac{\sigma^2}{8\pi^2B^2}f_3(u,v) + \frac{\sigma^2}{2N_{\rm vis}}\\
&\propto& u^2f_1(u,v) + v^2f_2(u,v) + 2uvf_3(u,v) + \frac{1}{2N_{\rm vis}},\label{error_vis3}
\end{eqnarray}
\normalsize
where $f_1,f_2,f_3$ are functions of the baseline lengths (array geometry). The residual error in each visibility due to each calibrator is \textit{independent} of the calibrator signal strength. Therefore, the impact on the visibilities of subtracting each calibrator has the same magnitude for all calibrators [but different distributions across visibilities; the covariances also depend on the source positions, ($l_i,m_i$)]. This result deviates from the assumptions of previous work, which assumed that the source position error was independent of signal strength, and therefore that residual error was greatest for the strongest calibrators \cite{datta10}.

\section{Impact on EoR statistical estimation}
One of the primary tools of a statistical measurement of the EoR is the angular power spectrum, which quantifies the signal power on a given angular scale. The angular power spectrum is defined by \cite{morales04,mcquinn06,bowman09,datta10}:
\begin{equation}
C_l = \frac{\displaystyle\sum_{(uv) \in l}N_{uv}|V_{uv}|^2}{\displaystyle\sum_{(uv) \in l}N_{uv}},
\end{equation}
where $N_{uv}$ is the number of visibilities contributing to a given ($uv$) cell, and the sum is over the ($uv$)-cells contributing to that $l$-mode ($l=2\pi{|\boldsymbol{u}|}$).

\begin{figure}
\begin{center}
\includegraphics[scale=0.6]{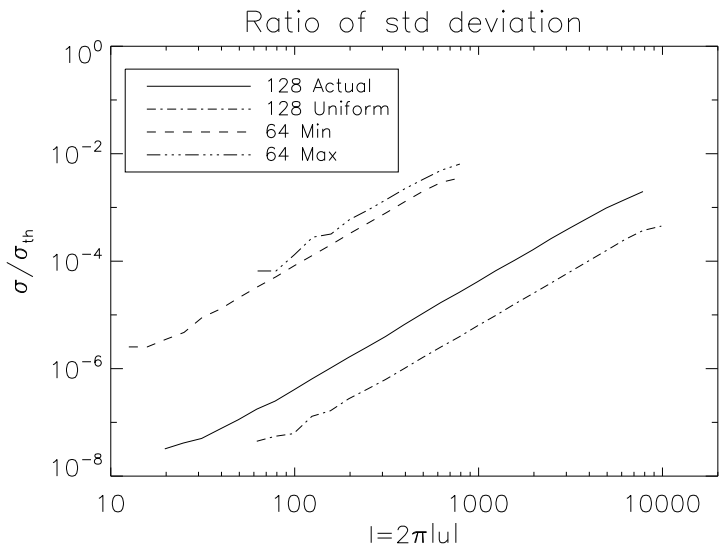}
\end{center}
\caption{Ratio of uncertainties in power for residual signal to thermal noise.}
\label{1d_variance}
\end{figure}

We propagate the uncertainties in the visibilities due to the thermal noise and residual point source signal to the angular power spectrum.
Figure \ref{1d_variance} displays the ratio of uncertainty in residual signal to thermal noise power (${\sigma_{C_l}}/{\sigma_{C_{l,{\rm therm}}}}$), for the four configurations.
In all cases the uncertainty in power exceeds the power from the point source subtraction. One can see the higher angular modes suffer more severely from point source subtraction, consistent with the additional noise in the long baseline visibilities (equation \ref{error_vis3}). These figures also highlight the differences between the array configurations; the longer baselines of the MWA sample higher $l$-modes, and the larger number of receiving elements yield improved sensitivity. Conversely, the larger instantaneous bandwidth of the PAPER array improves point source precision, offsetting the sensitivity degradation. The larger field-of-view of PAPER corresponds to a larger number of peeled sources, increasing the contribution from the residual signal compared with the thermal noise.

\section{Optimal information balance: data versus priors}
Figure \ref{hcrlb_precision}(a) demonstrates that the measurable sky position precision can exceed one arcminute for weak sources and low-sensitivity arrays. In this regime, the ionospheric refraction on short timescales (8--10 seconds) may be comparable to, or smaller than, the precision available from the dataset alone. An optimal estimator would balance the information available from the current dataset and previous timesteps to obtain the most accurate and precise position estimate (accuracy refers to parameter bias, and one must be careful not to bias the estimate by improper use of prior information). Using the HCRLB formalism derived in Section \ref{hcrlb}, we compute the precision available with an optimal estimator for two example characteristic refraction scales (60", 10"), and display the results in Figure \ref{hcrlb_precision}(b, c).
It is evident that for weak signals, the prior information dominates the Fisher matrix, while estimation for bright sources relies solely on the information available in the current dataset. These results suggest that in conditions of low ionospheric activity, there are substantial advantages available by considering a hybrid estimation scheme.

\section{Summary}
Point source estimation on short timescales is a primary observing task for upcoming low-frequency instruments, both for fundamental calibration and removal of foregrounds for statistical EoR experiments. Precise and accurate estimation is therefore critical to the quality and fidelity of all downstream quantitative science results. In this work we considered the optimal estimation of bright point source sky positions, and studied the impact of limited dataset information on the noise level in visibilities, and on EoR angular power spectrum estimation. We found that the magnitude of the signal is expected to be below the thermal noise uncertainty, and therefore not a limiting factor for EoR estimation. We do find, however, that the signal is structured differently to the thermal signal (it is coloured), and should be considered in the data analysis.

Incorporation of prior information about point source position can improve the precision with which positions can be estimated. For weak sources and compact, low-sensitivity arrays, the characteristic ionospheric refraction scale may be comparable to the precision available from the data, and an optimal estimator would use a balance of prior and current information.


\begin{thebibliography}{99}
  \bibitem{bowman09}
J.~D. {Bowman}, M.~F. {Morales}, and J.~N. {Hewitt}.
\newblock \emph{Foreground Contamination in Interferometric Measurements of the
  Redshifted 21 cm Power Spectrum}.
\newblock {\em ApJ}, 695:183--199, 2009.

\bibitem{cohen09}
A.~S. {Cohen} and H.~J.~A. {R{\"o}ttgering}.
\newblock \emph{Probing Fine-Scale Ionospheric Structure with the Very Large Array
  Radio Telescope}.
\newblock {\em AJ}, 138:439--447, 2009.

\bibitem{datta10}
A.~{Datta}, J.~D. {Bowman}, and C.~L. {Carilli}.
\newblock \emph{Bright Source Subtraction Requirements for Redshifted 21 cm
  Measurements}.
\newblock {\em ApJ}, 724:526--538, 2010.

\bibitem{ellingson09}
S.~W. {Ellingson}, T.~E. {Clarke}, A.~{Cohen}, J.~{Craig}, N.~E. {Kassim},
  Y.~{Pihlstrom}, L.~J. {Rickard}, and G.~B. {Taylor}.
\newblock \emph{The Long Wavelength Array}.
\newblock {\em IEEE Proceedings}, 97:1421--1430, 2009.

\bibitem{kay93}
S.~M. {Kay}.
\newblock {\em Fundamentals of statistical signal processing: estimation
  theory}.
\newblock Prentice-Hall, New York 1993.

\bibitem{lonsdale09}
C.~J. {Lonsdale}, R.~J. {Cappallo}, M.~F. {Morales}, {et al.}.
\newblock \emph{The Murchison Widefield Array: Design Overview}.
\newblock {\em IEEE Proceedings}, 97:1497--1506, 2009.

\bibitem{mcquinn06}
M.~{McQuinn}, O.~{Zahn}, M.~{Zaldarriaga}, L.~{Hernquist}, and S.~R.
  {Furlanetto}.
\newblock \emph{Cosmological Parameter Estimation Using 21 cm Radiation from the
  Epoch of Reionization}.
\newblock {\em ApJ}, 653:815--834, 2006.

\bibitem{morales04}
M.~F. {Morales} and J.~{Hewitt}.
\newblock \emph{Toward Epoch of Reionization Measurements with Wide-Field Radio
  Observations}.
\newblock {\em ApJ}, 615:7--18, 2004.

\bibitem{parsons10}
A.~R. {Parsons}, D.~C. {Backer}, G.~S. {Foster}, {et al.}.
\newblock \emph{The Precision Array for Probing the Epoch of Re-ionization: Eight
  Station Results}.
\newblock {\em AJ}, 139:1468--1480, 2010.

\bibitem{stappers11}
B.~W. {Stappers}, J.~W.~T. {Hessels}, A.~{Alexov}, {et al.}.
\newblock \emph{Observing pulsars and fast transients with LOFAR}.
\newblock {\em A\&A}, 530:A80, 2011.

\bibitem{tingay12}
S.~J. {Tingay et al.}
\newblock \emph{The Murchison Widefield Array}.
\newblock {\em ApJ, in prep.}, 2012.

\bibitem{trott11}
C.~M. {Trott}, R.~B. {Wayth}, J.-P.~R. {Macquart}, and S.~J. {Tingay}.
\newblock \emph{Source Detection in Interferometric Visibility Data. I. Fundamental
  Estimation Limits}.
\newblock {\em ApJ}, 731:81--+, 2011.

\bibitem{wijnholds08}
S.~J. {Wijnholds} and A.-J. {van der Veen}.
\newblock \emph{Fundamental Imaging Limits of Radio Telescope Arrays}.
\newblock {\em IEEE Journal of Selected Topics in Signal Processing}, 2:613-623, 2008.

\end{thebibliography}

\end{document}